\documentclass[journal,transmag,twoside]{IEEEtran}

\usepackage{xcolor} 
\definecolor{ieeeBlue}{rgb}{0,0,1} 

\usepackage[
colorlinks=true,
linkcolor=ieeeBlue,   
citecolor=ieeeBlue,   
urlcolor=black,       
filecolor=ieeeBlue,   
pdfborder={0 0 0},    
bookmarks=true,       
hyperindex=true       
]{hyperref}

\usepackage{cite}
\usepackage{orcidlink}

\usepackage{amsmath}
\usepackage{esint}
\usepackage{mathrsfs}
\usepackage{siunitx}

\usepackage{booktabs}
\usepackage{multirow}
\usepackage{balance}

\usepackage{graphicx}

\ifCLASSOPTIONcompsoc
\usepackage[caption=false,font=normalsize,labelfont=sf,textfont=sf]{subfig}
\else
\usepackage[caption=false,font=footnotesize]{subfig}
\fi

\renewcommand{\vec}[1]{\mathbf{#1}}
\begin{document}
	
\title{Stray Losses in Power Transformers: A Comparative\\
		Study of Modeling Frameworks for Mild Steel}
	
\author{\IEEEauthorblockN{
			Micha Swaanen\IEEEauthorrefmark{1,2}\,\orcidlink{0009-0004-7863-0237},
			Timo Overboom\IEEEauthorrefmark{2}\,\orcidlink{0000-0003-3112-747X},
			Siamak Pourkeivannour\IEEEauthorrefmark{2}\,\orcidlink{0000-0002-4684-1054},
			Tom Huiskamp\IEEEauthorrefmark{1}\,\orcidlink{0000-0002-8450-2600}, and
			Guus Pemen\IEEEauthorrefmark{1}\,\orcidlink{0000-0001-7550-2361}
		}
		\IEEEauthorblockA{\IEEEauthorrefmark{1}Department of Electrical Engineering, Eindhoven University of Technology, Eindhoven, The Netherlands}
		\IEEEauthorblockA{\IEEEauthorrefmark{2}Royal SMIT Transformers B.V., Nijmegen, The Netherlands}%
		\thanks{Corresponding author: M. P. P. Swaanen (email: m.p.p.swaanen@student.tue.nl).}}
	
\markboth{PREPRINT SUBMITTED TO IEEE}%
{Swaanen \MakeLowercase{\textit{et al.}}: A Comparative Study of Modeling Frameworks for Mild Steel}
	
\IEEEtitleabstractindextext{%
\begin{abstract}
Accurately predicting eddy-current and hysteresis losses in mild steel components such as the structural parts and tank of power transformers remains a challenge in transformer design. Conventional numerical approaches for eddy-current loss analysis require discretization at a fraction of the skin depth. While accurate, such fine volumetric mesh becomes computationally prohibitive for large power transformers. Moreover, hysteresis loss is not always included in this analysis. To investigate the trade-off between computational cost and accuracy, this work presents a comparative analysis of eight distinct modeling frameworks for predicting both eddy-current and hysteresis losses in mild steel. These include efficient methods avoiding meshing, such as standard impedance boundary conditions, and more accurate approaches retaining full volumetric meshing, namely complex permeability, nonlinear single-valued $B{-}H$ curve, and Jiles-Atherton hysteresis modeling. To accurately represent the material's electromagnetic behavior, the properties of the mild steel used are measured directly from samples. Finally, loss measurements are conducted to benchmark all eight frameworks, demonstrating mean absolute percentage errors between predictions and measurements generally below 20\%.
\end{abstract}
		
}
	
\maketitle
\IEEEdisplaynontitleabstractindextext
\IEEEpeerreviewmaketitle
	

\section{Introduction}
\IEEEPARstart{P}{ower} transformers are essential components of modern power systems and will remain so for the foreseeable future. The generation of alternating-current power by synchronous generators necessitates voltage transformations across the power system to enable efficient transmission and distribution of electric power \cite{glover2011power, heathcote2007transformer}.

The continuous drive toward higher power and voltage ratings pushes modern transformers toward transportation and manufacturing limits \cite{fogelberg2007transforming, doe2014lpt}. To maximize reliability and cost-effectiveness over its 25-50 year lifetime, large power transformers are designed to achieve efficiencies as high as 99.7\% \cite{del2017transformer}. However, at power ratings in the hundreds of MVA, even fractional percentage losses dissipate tens to hundreds of kilowatts of power. This thermal burden causes mechanical stress and accelerates the degradation of both the insulation and transformer oil \cite{sv2004transformer, winders2002power, li2013}. Furthermore, as nominal capacities continue to rise, stray leakage fields intensify, exacerbating the issue of localized overheating (hotspots). To mitigate these hotspots and avoid premature failure or financial penalties, manufacturers like Royal SMIT Transformers must accurately predict all losses during the initial design stage.

\IEEEpubidadjcol In a transformer, load losses consist of DC and eddy-current winding losses, alongside stray losses in other parts of the transformer such as the clamping and tank. These stray losses are induced by leakage magnetic fields from the windings that interact with structural steel, producing both eddy-current and hysteresis losses \cite{Piccoli2018}. While methods for designing active transformer parts (cores and windings) and calculating their internal losses are well established \cite{sv2004transformer}, accurately predicting stray losses in inactive structural components remains a challenge \cite{johnson2014simulation}. Consequently, these structural stray losses are the primary focus of this work.

Stray losses generally account for 5–20\% of the total load losses, but can reach as high as 34\% \cite{sv2004transformer, mikuli2026}. They are directly influenced by geometry, magnetic field distribution, frequency and material properties and processing \cite{sv2004transformer}. Moreover, because each transformer is custom-engineered to meet specific client requirements, a thorough analysis is required for every manufactured unit. In the context of stray loss analysis, loss calculations increasingly rely on computational techniques such as the finite element method (FEM). The accuracy of these numerical computations is directly dependent on the chosen discretization size and modeling framework. Due to their high permeability and conductivity, transformer mild steel components, such as the clamping structure and tank, typically have a skin depth that is far smaller than their physical dimensions (approximately 2\,\si{mm} for saturated mild steel at 50\,\si{Hz} \cite{stoll1974analysis}). Consequently, meshing at the skin-depth scale imposes a substantial and potentially prohibitive computational cost.

Transformer stray loss predictions often assume a constant, real relative permeability for structural steel like mild steel \cite{sv2004transformer, Kulkarni2013, del2017transformer}. While this assumption facilitates comparison with analytical calculations, it inherently neglects magnetic saturation. Furthermore, these models compute only eddy-current loss, omitting the hysteresis contribution. Several studies have investigated eddy-current loss distributions in steel plates and compared them with measurements using the TEAM benchmark Problem 21 \cite{Takahashi2001, Cheng2004, Zhiguang2006, Cheng2009}. Due to the physical complexity of magnetic hysteresis, its effects are usually incorporated through approximations, such as post-processing techniques or the application of a complex permeability that models the hysteresis loop as an ellipse. The latter approach has been extensively researched, though predominantly for ferrites and electrical steels \cite{hollaus2002, o1972flux, Strac2015}. For instance, \cite{Strac2015} utilizes a complex permeability to calculate losses in structural steels; however, the permeability is assumed constant and thus independent of the magnetic flux density. Others simplify meshing, and implement standard impedance boundary conditions to decrease computational burden, but also neglect hysteresis \cite{Duc2016Loss, Holland1992}.
In practice, true hysteresis loops exhibit a sigmoidal shape. To replicate this nonlinear behavior, phenomenological frameworks such as the Jiles-Atherton model are commonly employed \cite{jiles1986theory, szewczyk2014computational}.

Accordingly, this paper presents a comparative analysis of eight loss-modeling frameworks for mild steel structural components in transformers. Section~\ref{sec:propsteel} experimentally characterizes the electromagnetic properties of S355 mild steel required for numerical modeling. Section~\ref{sec:aclossshell} formulates the governing equations and constitutive relations on a benchmark ferromagnetic cylindrical shell. Section~\ref{sec:frameworksfea} introduces eight loss-modeling frameworks spanning various constitutive relations, solver domains, and discretization strategies. Section~\ref{sec:experimentbenchmark} validates these frameworks against measurements on a physical shell setup to benchmark their accuracy and computational cost. Finally, Section~\ref{sec:casestudy} evaluates all eight frameworks on a scale-model transformer geometry.

\section{Properties of mild steel}
\label{sec:propsteel}
This section outlines the structural components of medium and large power transformers and discusses the properties of the mild steel used in their fabrication. Additionally, it presents the electromagnetic properties acquired via measurements.

The clamping structure's design is primarily dictated by short-circuit forces, which can reach 100–900 times nominal levels \cite{del2017transformer}. Although intended for mechanical support, these highly permeable and conductive mild steel components influence the stray magnetic field distribution and loss densities.

\begin{figure}[!t]
	\centering
	\includegraphics[width=0.75\columnwidth]{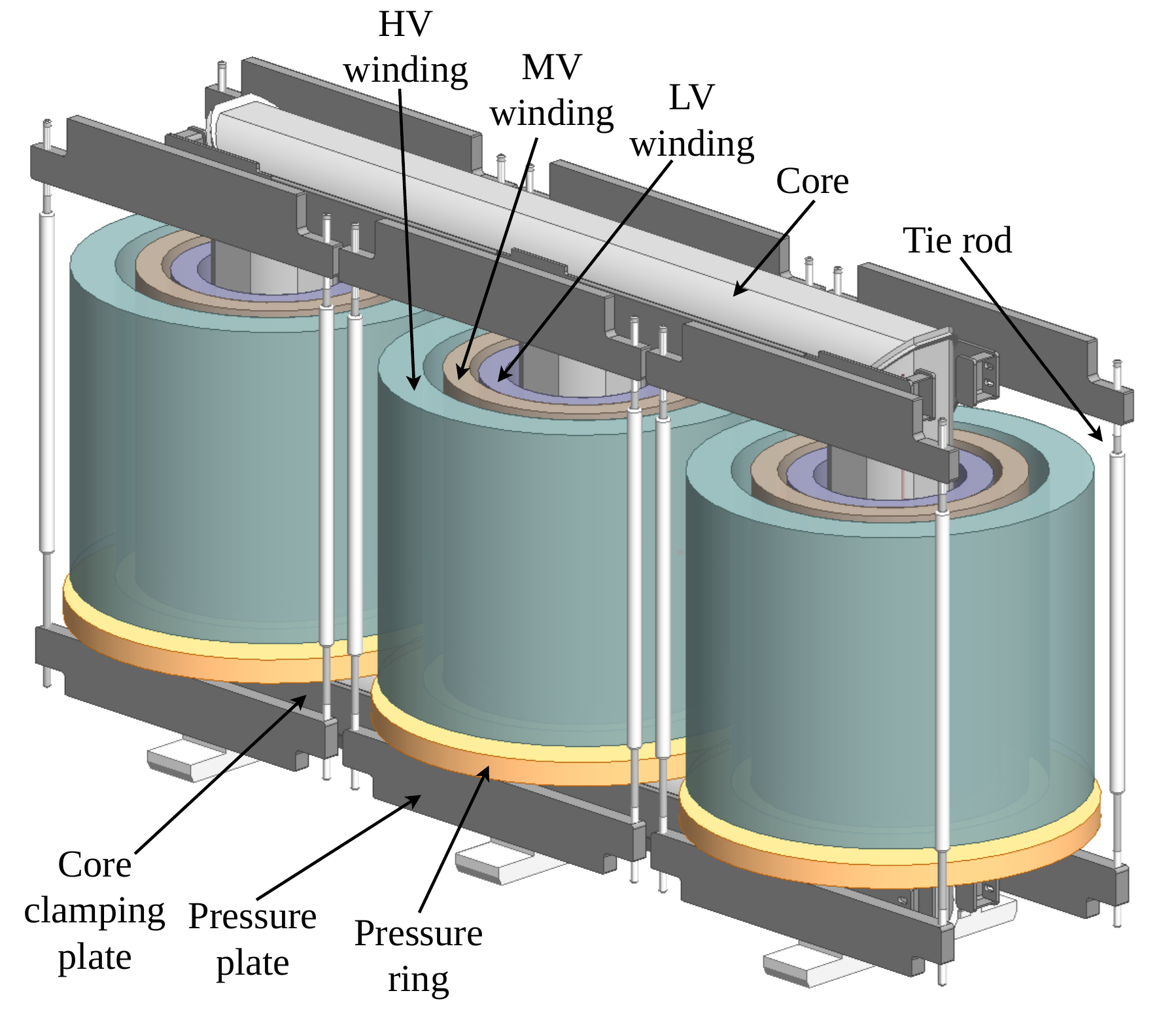}
	\caption{The active part of a typical medium-to-large power transformer, with mild steel components shown in dark grey (the tank is not included).} 
	\label{fig:trafo}
\end{figure}

Within this structure, core and winding clamping plates compress the core yokes and windings, respectively. The winding clamping plates are vertically connected by bolted tie plates or tie rods, providing compressive force to the windings via wooden key spacers. To prevent axial movement during short circuits, upper and lower pressure rings apply compressive force across each concentric set of high-, medium- and low-voltage windings. Together with the core, this assembly forms the active part and is housed within a mild steel tank. The tank contains insulating oil, acts as an environmental barrier, and partially shields against stray magnetic flux. Fig.~\ref{fig:trafo} depicts the active part of a typical power transformer manufactured by Royal SMIT Transformers.

Mild steels, also referred to as low-carbon steels, are used in many different structural applications. While technically soft magnetic materials, they exhibit coercivities approaching the $1\,\si{kA/m}$ threshold of hard magnetic materials \cite{IEC60404-1}. Transformers produced by Royal SMIT mostly use structural components made from grade S355, which has a yield strength of 355MPa. For this grade, and many other grades, extensive delivery conditions are specified by standard EN10025‑2 \cite{EN10025-2:2019}. It is worth noting that, unlike electrical steels used in transformer cores, these steels are not specifically designed for magnetic applications. Their electromagnetic behavior depends on composition, microstructure, and processing history \cite{Rumiche2008}, which emphasizes the need for methods that more accurately model them. Consequently, the material's electromagnetic properties must be experimentally quantified, which follows now.
\begin{table}[!t]
	\caption{Measured and average resistance values, as well as the corresponding conductivity, of two S355 samples.}
	\centering
	\begin{tabular}{@{}lcc@{}}
		\toprule
		Sample & Measured resistance (\si{\mu\Omega}) & Conductivity (\si{S/m}) \\ \midrule
		\multirow{2}{*}{Sample 1} & $1050$ & $5.35 \times 10^6$ \\
		& $1025$ & $5.48 \times 10^6$ \\ \addlinespace
		\multirow{2}{*}{Sample 2} & $1020$ & $5.51 \times 10^6$ \\
		& $1017$ & $ 5.52 \times 10^6$ \\ \midrule
		Sample average & $1028$ & $\sigma_{\mathrm{s}} = 5.465 \times 10^6$ \\ \bottomrule
	\end{tabular}
	\label{tab:s355_conductivity}
\end{table}%
\begin{figure}[!t]
	\centering
	\includegraphics[width=\columnwidth]{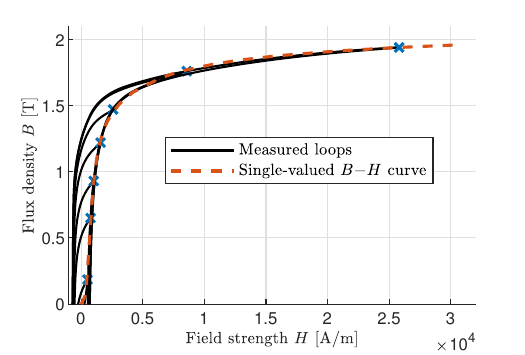}
	\caption{Seven hysteresis loops from a total of 45 measured loops, obtained via the DC hysteresis graph method. The positive extrema of each loop are marked and used to fit the single-valued $B{-}H$ curve.} 
	\label{fig:single_valued}
\end{figure}

To eliminate parasitic lead and contact resistances in low-resistance measurements, the electrical resistivity of two S355 ring samples is measured via four-terminal sensing using a LoRe precision micro-ohmmeter. The resulting S355 conductivity $\sigma_{\mathrm{s}}$ is presented in Table~\ref{tab:s355_conductivity}, and aligns with conductivity values presented in \cite{Strac2015}.

Hysteresis measurements are performed using the DC hysteresis graph method, standardized in IEC 60404-4 \cite{IEC60404-4}. In this method, the magnetic field strength and flux density are logged while applying a controlled, slowly varying excitation to a sample. The result is a record of a full hysteresis loop. This procedure is repeated $45$ times with amplitudes ranging from $300\,\si{A/m}$ to $56.2\,\si{kA/m}$ to capture multiple minor loops and the major loop. The quasi-static excitation keeps induced eddy currents negligible. Both the hysteresis energy density, as a function of peak flux density, and the single-valued $B{-}H$ curve are derived from the measured collection of loops. The hysteresis energy density, $W_\mathrm{h}$, is calculated by integrating the area of each loop individually via
\begin{equation}
	W_\mathrm{h} = \oint H \,\mathrm{d}B = \int_{0}^{T} H(t)\frac{\mathrm{d}B}{\mathrm{d}t}\,\mathrm{d}t,
	\label{eq:Whys}
\end{equation}
which is tabulated against the peak flux density of the corresponding loop. The single-valued nonlinear $B{-}H$ curve is obtained by fitting the following expression to the first-quadrant extrema of the measured loops:
\begin{equation}
	\label{eq:BHfitted}
	B(H) = \frac{c_1H + c_2H^2 + B_\mathrm{max}c_3H^3}{c_4 + c_5H + c_6H^2 + c_3H^3} + \mu_0H.
\end{equation}
Fig.~\ref{fig:Wh_meas_Stein_JA} shows hysteresis energy density and Fig.~\ref{fig:single_valued} plots the fitted single-valued $B{-}H$ curve alongside seven minor loops within the first two quadrants. Table~\ref{tab:BHStein_coef} (Appendix) lists the coefficients of \eqref{eq:BHfitted}. All subsequent modeling builds on these characterized electromagnetic properties.

\section{AC loss in a cylindrical magnetic shell}
\label{sec:aclossshell}
\begin{figure}[!t]
	\centering
	\includegraphics[width=0.7\columnwidth]{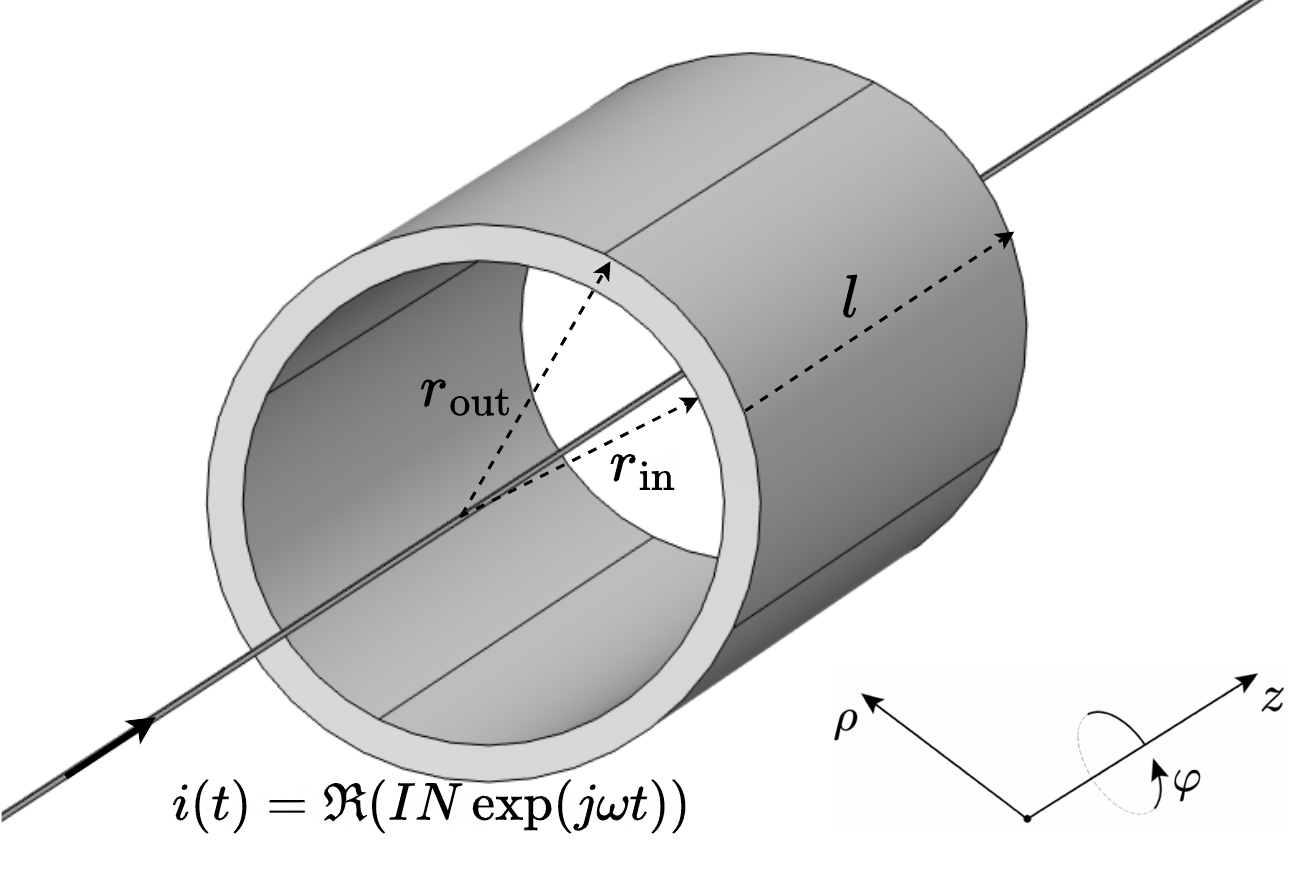}
	\caption{Geometry as considered in Section~\ref{sec:aclossshell} and during benchmarking all eight loss frameworks.} 
	\label{fig:cylinder_geom}
\end{figure}%
This section establishes the quasi-static Maxwell's equations and the constitutive relations that link the electromagnetic fields to the material properties. Applying these to a cylindrical magnetic shell (Fig.~\ref{fig:cylinder_geom}) yields analytical expressions for the electromagnetic fields, current density, and both eddy-current and hysteresis losses. The cylindrical geometry reduces to a one-dimensional electromagnetic problem, simplifying the analytical derivations. The shell features an inner radius $r_\mathrm{in}$, outer radius $r_\mathrm{out}$, conductivity $\sigma_{\mathrm{s}}$, and relative permeability $\mu_\mathrm{r}$. It encloses a central winding of $N$ uniformly distributed turns subjected to a sinusoidal current of amplitude $I$ and frequency $\omega = 2\pi f$. For analytical purposes, this winding is represented as an equivalent single centered conductor carrying a total current of $i(t) = \Re\{IN \exp(j\omega t)\}$.
\subsection{Derivation of analytic equations}
In the low-frequency regime considered, the displacement current can be safely neglected, yielding the quasi-static approximation of Maxwell's equations. The governing field equations are dictated by Ampère's law and Faraday's law of induction, given respectively by
\begin{subequations}
	\label{eq:maxwell}
	\begin{align}
		\nabla \times \vec{H} &= \vec{J},
		\label{eq:ampere} \\
		\nabla \times \vec{E} &= -\frac{\partial \vec{B}}{\partial t},
		\label{eq:faraday}
	\end{align}
\end{subequations}
These aforementioned field quantities are coupled through two primary constitutive relations, which provide the essential link via the properties of the specific medium. These are
\begin{subequations}
	\label{eq:constitutive}
	\begin{align}
		\vec{B} &= \mu \vec{H},
		\label{eq:const_B} \\
		\vec{J} &= \sigma \vec{E},
		\label{eq:const_J}
	\end{align}
\end{subequations}
where $\mu = \mu_0 \mu_\mathrm{r}$ is the magnetic permeability and $\sigma$ is the electrical conductivity. 

While the formulation in \eqref{eq:const_B} describes the magnetic behavior through a single lumped permeability parameter $\mu$, nonlinear ferromagnetic modeling often requires explicitly separating the applied magnetic field from the material's magnetization. This is achieved by expressing the magnetic flux density via the Sommerfeld convention as
\begin{equation}
	\vec{B} = \mu_0 \left( \vec{H} + \vec{M} \right),
	\label{eq:sommerfeld}
\end{equation}
where $\mu_0$ is the permeability of free space and $\vec{M}$ is the magnetization field. For a linear, isotropic medium, the magnetization is proportionally linked to the magnetic field strength via the magnetic susceptibility $\chi$, such that $\vec{M} = \chi \vec{H}$. Substituting this relationship back into the Sommerfeld convention yields
\begin{equation}
	\vec{B} = \mu_0 \left( 1 + \chi \right) \vec{H},
\end{equation}
which reconciles with \eqref{eq:const_B} by defining the relative permeability as $\mu_\mathrm{r} = 1 + \chi$. 

By taking the curl of Ampère's law and substituting both Faraday's law and Ohm's law from \eqref{eq:const_J}, the interdependent field equations can be decoupled. Assuming a homogeneous, isotropic medium in the time-harmonic domain, this manipulation leads to the well-known Helmholtz as
\begin{equation}
	\nabla^2 \vec{H} - \alpha^2 \vec{H} = 0.
	\label{eq:helmholtz}
\end{equation}
The instantaneous magnetic and electric fields for the shell in cylindrical coordinates follow from the right-hand rule as
\begin{subequations}
	\begin{align}
		&\vec{H}(\rho,t) = \Re{\left(H(\rho)\exp{(j\omega t)} \vec{a}_\varphi\right)}, \\ 
		&\vec{E}(\rho,t) = \Re{\left(E(\rho)\exp{(j\omega t)} \vec{a}_z\right)},
	\end{align}
\end{subequations}
where $H(\rho)$ and $E(\rho)$ are the scalar complex amplitudes. Solving the Helmholtz equation yields these amplitudes in terms of modified Bessel functions of the first kind, $I_1(\cdot)$, and the second kind, $K_1(\cdot)$ \cite{jeffrey1995handbook}:
\begin{subequations}
	\begin{align}
		H(\rho) &= C_1 I_1(\alpha\rho) + C_2 K_1(\alpha\rho), \\
		E(\rho) &= \frac{\alpha}{\sigma_{\mathrm{s}}} \left( C_1 I_0(\alpha\rho) - C_2 K_0(\alpha\rho) \right),
	\end{align}
\end{subequations}
where $\alpha = \sqrt{j\omega\sigma_{\mathrm{s}}\mu_0\mu_\mathrm{r}}$. The integration constants $C_1$ and $C_2$ are evaluated by applying Ampère's law at the shell's inner and outer boundaries, yielding the Dirichlet conditions $H(r_\mathrm{in}) = \frac{I}{2\pi r_\mathrm{in}}$ and $H(r_\mathrm{out}) = \frac{I}{2\pi r_\mathrm{out}}$.

The time-average total power loss $P_\mathrm{tot}$ per unit length is evaluated via the complex Poynting vector. Because the flux vanishes on the end caps, integrating the purely radial flux over the inner and outer boundaries simplifies directly to
\begin{equation}
	P_\mathrm{tot}=\pi\Re(r_\mathrm{out}E(r_\mathrm{out})H^*(r_\mathrm{out})-r_\mathrm{in}E(r_\mathrm{in})H^*(r_\mathrm{in})).
\end{equation}
\begin{table}[!t]
	\caption{Parameters considered in the cylindrical shell geometry during analytical and FEA simulation treatment.}
	\centering
	\begin{tabular}{@{}llll@{}}
		\toprule
		Quantity            & Symbol    & Value  & Unit     \\ \midrule
		Inner shell radius   & $r_\mathrm{in}$    & $66.2$    &\si{mm}      \\
		Outer shell radius & $r_\mathrm{out}$ & $76.2$ &\si{mm} \\
		Current angular frequency & $\omega$  &  $2\pi60$  &\si{rad/s} \\
		Peak current amplitude & $I$ & $35\sqrt{2}$ &\si{A} \\
		Number of turns & $N$ & $107$ & $-$ \\
		Conductivity of the shell & $\sigma_{\mathrm{s}}$ & $5.465$ &\si{MS/m} \\
		Relative permeability of the shell & $\mu_\mathrm{r}$ & see Fig.~\ref{fig:cylinder_loss} & $-$ \\
		Phase lag between $\vec{B}$ and $\vec{H}$ & $\theta$ & see Fig.~\ref{fig:cylinder_loss} &$^\circ$ \\
		\bottomrule
	\end{tabular}
	\label{tab:cylinder_parameters}
\end{table}%
The aforementioned total loss comprises eddy-current and hysteresis components. The time-averaged eddy-current loss per unit length, $P_\mathrm{e}$, is obtained by integrating the eddy-current loss density over the cylindrical cross-section, yielding
\begin{equation}
	\label{eq:Pe}
	P_\mathrm{e}=\frac{\pi}{\sigma_{\mathrm{s}}}
	\int_{r_\mathrm{in}}^{r_\mathrm{out}} J(\rho)J^*(\rho)\rho\,d\rho.
\end{equation}
The second component, the hysteresis loss, can be included by introducing a phase lag $\theta$ throughout the material such that $\vec{B}$ lags $\vec{H}$. This parameter is commonly referred to in the literature as the hysteresis angle \cite{sv2004transformer, stoll1974analysis}. The angle is integrated into the derived model by replacing the real relative permeability $\mu_\mathrm{r}$ in \eqref{eq:const_B} with a complex valued one like 
\begin{equation}
	\mu_\mathrm{r}=|\mu_\mathrm{r}|\exp(-j\theta).
	\label{eq:complexper}
\end{equation}
Consequently, the hysteresis loss per unit length $P_\mathrm{h}$ is
\begin{equation}
	P_\mathrm{h} = P_\mathrm{tot} - P_\mathrm{e}.
	\label{eq:Ph}
\end{equation}

Equations \eqref{eq:Pe} and \eqref{eq:Ph} enable an analysis of how material properties like the conductivity, relative permeability, and hysteresis angle affect both loss components. Beside being calculated analytically, these values were also computed in FEMM \cite{meeker_femm}. The parameters listed in Table~\ref{tab:cylinder_parameters} are selected not only for comparison purposes here, but also to reflect the physical experimental setup described in Section~\ref{sec:experimentbenchmark}. A parametric sweep of $\theta$ is performed for $\mu_\mathrm{r} \in \{200, 1000\}$.

\begin{figure}[!t]
	\centering
	\includegraphics[width=\columnwidth]{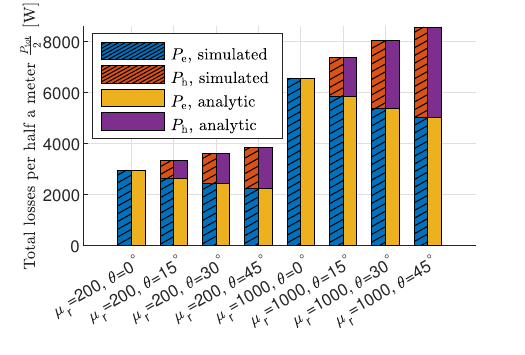}
	\caption{Comparison of eddy-current and hysteresis losses in the cylindrical shell: FEA simulations versus the analytical equations from Section~\ref{sec:aclossshell}. Losses are expressed per half a meter, and other parameters are as in Table~\ref{tab:cylinder_parameters}.} 
	\label{fig:cylinder_loss}
\end{figure}%

Fig.~\ref{fig:cylinder_loss} demonstrates agreement between the simulations and analytical calculations. More importantly, the results highlight that neglecting the hysteresis effect leads to a significant underestimation of total losses, as well as a high dependence on relative permeability. For transformer components fabricated from mild steel, analytical approaches for calculating losses become intractable because of their complex 3D geometries. Instead, accurate loss evaluation is typically achieved via 3D FEM analysis. The accuracy and computational cost depend on the specific frameworks utilized to model the material's magnetic behavior and losses, which are discussed now.

\section{Loss frameworks for FEA}
\label{sec:frameworksfea}
\begin{figure}[!t]
	\centering
	\includegraphics[width=\columnwidth]{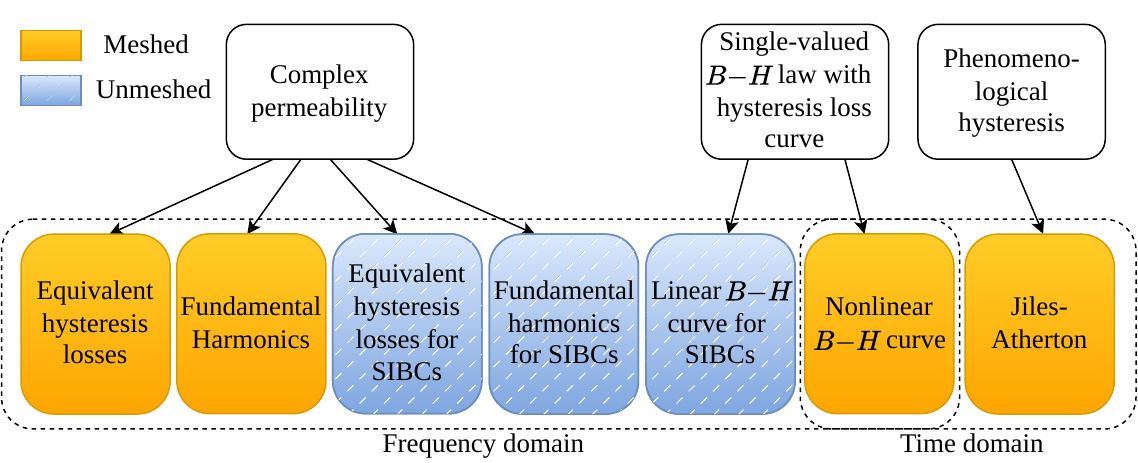}
	\caption{All loss frameworks considered in this work, including their study type and mesh category (meshed or unmeshed).} 
	\label{fig:frameworks}
\end{figure}

This section details eight modeling frameworks for computing eddy-current and hysteresis losses in bulk mild steel (Fig.~\ref{fig:frameworks}). Excess losses are omitted but are detailed in \cite{zeinali2023magnetodynamic}. To systematically compare these approaches, the frameworks are categorized by two fundamental aspects: their assumed constitutive relationships and their finite-element implementations.

First, the frameworks are distinguished by their constitutive relationship, which defines how the magnetic behavior of the mild steel is approximated. As shown in Fig.~\ref{fig:frameworks}, the material is characterized using one of three relationships: a complex permeability using \eqref{eq:const_B}, a single-valued $B{-}H$ law using \eqref{eq:const_B} (where the permeability depends on the field strength), or a phenomenological hysteresis model using \eqref{eq:sommerfeld}.

Second, the frameworks are categorized by their finite-element implementation, which encompass both the solver domain and the spatial discretization strategy. As shown at the bottom of Fig.~\ref{fig:frameworks}, the governing equations are solved either in the frequency domain or the time domain. Furthermore, the discretization of the conductive regions varies; the material is either meshed or unmeshed by applying a standard impedance boundary condition (SIBC). These concepts of solver-domain selection and discretization requirements are briefly discussed.

The frequency-domain formulation assumes time-harmonic field behavior, solving the governing equations for phasors rather than time-varying fields. This approach is computationally efficient, as time derivatives correspond to multiplications by $j\omega$. However, this formulation is restricted to linear media or nonlinear materials using an effective constitutive relation. Conversely, time-domain analysis solves time-varying fields via sequential time-stepping, which is essential for capturing history-dependent material nonlinearities, such as hysteresis modeled by Jiles-Atherton.

The conventional approach to evaluating electromagnetic fields and their associated losses in magnetic materials relies on full volumetric discretization of the domain using the finite element method. This technique explicitly solves for the electromagnetic fields across the entire mesh to compute volumetric loss densities. Accurate resolution of eddy currents typically requires at least three to four mesh elements across the skin depth, which can become challenging in large geometries containing highly permeable and conductive materials. Dense meshing generates an excessively large reluctivity matrix, inflating memory requirements and computation times, particularly for three-dimensional problems.

\begin{figure}[!t]
	\centering
	\includegraphics[
	trim={0cm 1.95cm 0cm 1.45cm},
	clip, width=0.94\columnwidth]{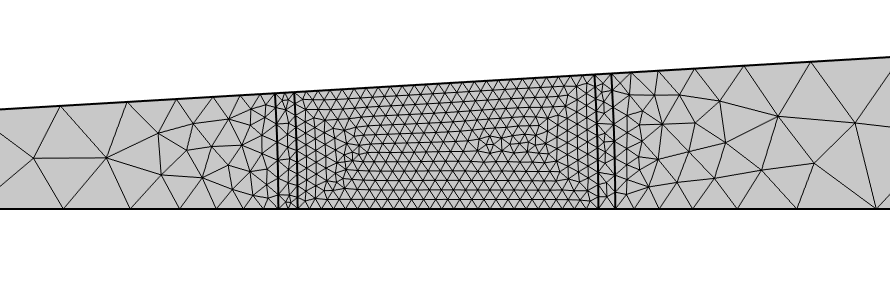}%
	\\
	\includegraphics[
	trim={0cm 0.8cm 0cm 1.45cm},
	clip, width=0.94\columnwidth]{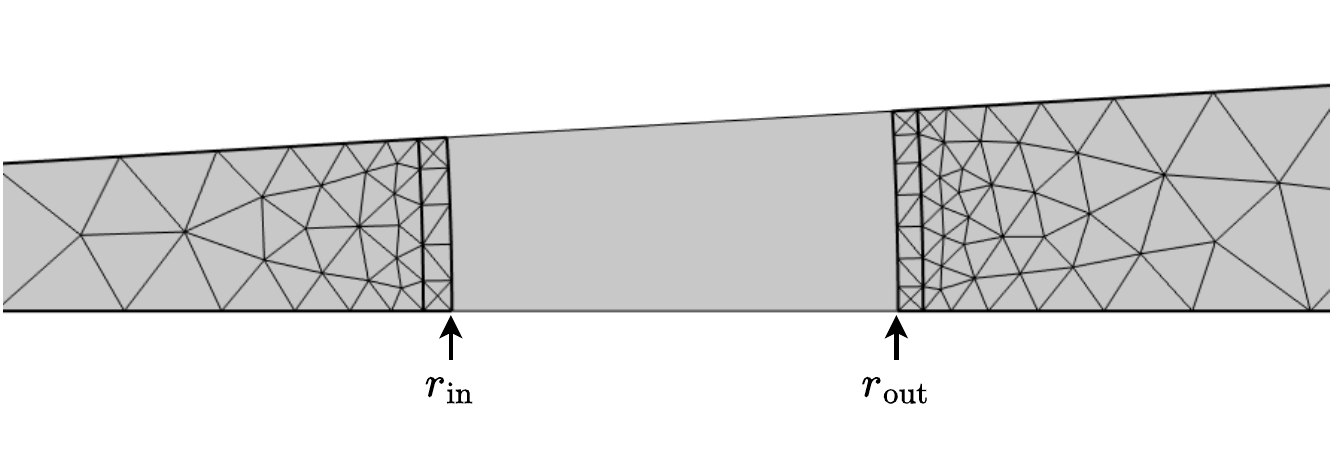}
	\caption{Meshing for a permeable, conductive cylindrical shell, comparing a fully meshed steel domain (top) with an unmeshed domain utilizing two SIBCs (bottom).}
	\label{fig:shell_mesh}
\end{figure}

Alternatively, the standard impedance boundary condition, is particularly effective when the skin depth is small relative to the material thickness \cite{senior1995approximate, preston1983three}. Applying an SIBC eliminates the need to mesh the interior domain of the material. Fig.~\ref{fig:shell_mesh} illustrates this difference in meshing for the cylindrical shell. The SIBC approximates the steel domain as a linear surface coupling boundary, relating the tangential electric field and magnetic field strength via the surface impedance, $Z_\mathrm{s}$:
\begin{equation}
	\vec{n} \times (\vec{n} \times \vec{E}) = -Z_\mathrm{s} (\vec{n} \times \vec{H}),
\end{equation}
where $\vec{n}$ is the surface normal vector and points into the steel. For a conductive material ($\sigma \gg \omega\epsilon$), $Z_\mathrm{s}$ reduces to
\begin{equation}
	Z_\mathrm{s} =\frac{\mu\omega}{\sqrt{-j\omega\mu\sigma(1+j\frac{\omega\epsilon}{\sigma})}}\approx\sqrt{\frac{j\omega\mu}{\sigma}},
\end{equation}
where $\mu$ and $\sigma$ are the permeability and conductivity of the material, respectively. A critical assumption in this work is that the steel thickness is much greater than the skin depth.

For the 2D FEA problem, the shell must be modeled as an electrically floating conductor, to prevent net circulating current. Accordingly, the zero net surface current constraint is applied via the out-of-plane tangential surface electric field
\begin{equation}
	\oint_{\mathscr{C}_1} |\vec{E_\mathrm{s}}| \, dl + \oint_{\mathscr{C}_2} |\vec{E_\mathrm{s}}| \, dl = 0,
\end{equation}
where $\mathscr{C}_1$ and $\mathscr{C}_2$ denote the inner and outer shell surface contours. $|\vec{E_\mathrm{s}}|$ is the surface electric field amplitude, and $dl$ is the differential arc length along the respective contour. Due to the inherent continuous geometry of 3D FEA models, explicit enforcement of such conditions is not required.

Note that commercial software suites may not directly support specific constitutive relationships, such as the Jiles-Atherton model and flux-dependent hysteresis angle, or certain discretization strategies like SIBC.

\subsection{Complex permeability}
In the frequency domain, hysteresis is typically modeled using a complex permeability that approximates the measured hysteresis loop as an equivalent ellipse. The constitutive relationship is given by \eqref{eq:const_B}, and the complex permeability by \eqref{eq:complexper} \cite{hollaus2002, o1972flux}. In contrast to Section~\ref{sec:aclossshell}, both the permeability modulus $\mu_\mathrm{r}$ and the hysteresis angle $\theta$ depend on the magnetic flux density. This section introduces two methods to derive this dependence, which are then extended using SIBCs to avoid meshing the steel domain.

\subsubsection{Equivalent hysteresis losses}
\label{subsubsec:EHL}
\begin{figure}[!t]
	\centering
	\includegraphics[width=\columnwidth]{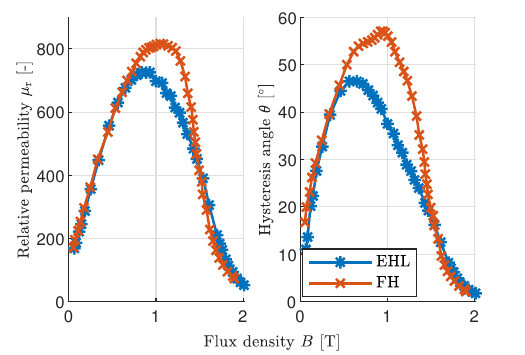}
	\caption{The complex permeability modulus and hysteresis angle for S355 mild steel by the EHL and FH method, as functions of the local magnetic flux density amplitude.} 
	\label{fig:murthetaS355}
\end{figure}

The equivalent hysteresis loss (EHL) method determines the equivalent ellipse by matching the measured peak field values, $B_\mathrm{p}$ and $H_\mathrm{p}$, while preserving the per-cycle hysteresis loss. Matching the peak values directly yields the relative permeability modulus:
\begin{equation}
	\mu_\mathrm{r} = \frac{B_\mathrm{p}}{\mu_0 H_\mathrm{p}}.
	\label{eq:murBpHp}
\end{equation}
To preserve the loss, the hysteresis angle $\theta$ is derived by equating the area of the equivalent ellipse to the measured hysteresis energy density, $W_\mathrm{h}$. Substitution of sinusoidal expressions for both the magnetic field strength and the flux density, separated by a phase angle $\theta$, into \eqref{eq:Whys} yields the following:
\begin{equation}
	\theta = \arcsin\left(\frac{\mu_0\mu_\mathrm{r} W_\mathrm{h}}{\pi B_\mathrm{p}^2}\right).
\end{equation}
The calculated results for S355 steel are shown in Fig.~\ref{fig:murthetaS355}. Moreover, Fig.~\ref{fig:twoloops} plots the approximations generated using the permeability modulus and hysteresis angle for two loops.

\begin{figure}[!t]
	\centering
	\includegraphics[width=\columnwidth]{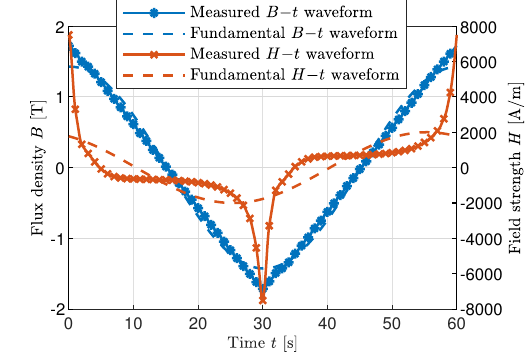}
	\caption{The measured and fundamental time-domain waveforms of the field strength and flux density, obtained via the DC hysteresis graph method.} 
	\label{eq:timefield}
\end{figure}

\subsubsection{Fundamental harmonics}
\begin{figure}[!t]
	\centering
	\includegraphics[width=\columnwidth]{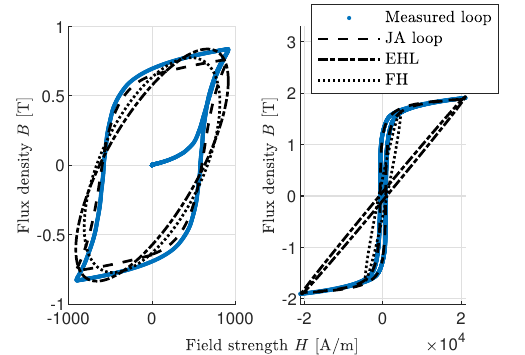}
	\caption{Two measured hysteresis loops, alongside their elliptic magnetization curve approximations (via the EHL and FH methods) and the Jiles-Atherton simulated curve.} 
	\label{fig:twoloops}
\end{figure}

An alternative to the EHL method is the fundamental harmonics (FH) method. The FH method calculates the complex permeability from the fundamental harmonic of the measured magnetic field strength and flux density time waveforms. A Fourier transform is applied to both waveforms to extract the fundamental amplitudes, $H_1$ and $B_1$, and phases, $\phi_H$ and $\phi_B$, of the magnetic field strength and flux density, respectively. The result of this procedure for one loop is shown in Fig.~\ref{eq:timefield}. The phase lag (hysteresis angle) between the two fundamental components is evident, alongside the saturation of the material. Subsequently, the relative permeability modulus is calculated using both fundamental amplitudes:
\begin{equation}
	\mu_\mathrm{r} = \frac{B_1}{\mu_0 H_{1}}.
\end{equation}
The hysteresis angle is computed by the phase difference:
\begin{equation}
	\theta = \phi_H - \phi_B.
\end{equation}
The results for S355 steel are shown in Fig.~\ref{fig:murthetaS355}. Moreover, Fig.~\ref{fig:twoloops} plots the approximations generated using the permeability modulus and hysteresis angle for two loops. Saturation yields a low relative permeability and hysteresis angle at high flux densities, while both quantities maximize around the knee of the magnetization curve, for both the EHL and FH method.

\subsubsection{Equivalent hysteresis losses and fundamental harmonics for SIBCs}
\begin{figure}[!t]
	\centering
	\includegraphics[
	trim={0cm 0.6cm 0cm 0.3cm},
	clip,width=0.85\columnwidth]{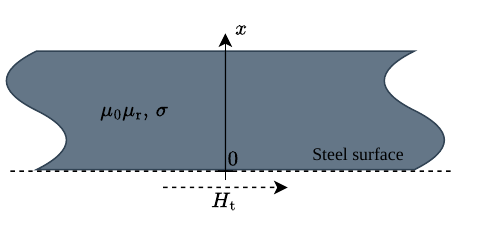}
	\caption{Semi-infinite linear steel half-space, occupying $x>0$, subjected to a tangential magnetic field strength $H_\mathrm{t}$.} 
	\label{fig:Hs_interface}
\end{figure}

To implement SIBCs based on the EHL and FH frameworks, the steel must be modeled as a magnetically linear conducting medium. To establish the required equivalent material properties used within the impedance boundary conditions, a semi-infinite steel half-space occupying $x>0$ is analyzed under a tangential magnetic field strength $H_\mathrm{t}$, as illustrated in Fig.~\ref{fig:Hs_interface}. The eddy-current and hysteresis losses within the half-space are characterized by an equivalent relative permeability modulus, $\mu_\mathrm{r}^\mathrm{eq}$, and an equivalent hysteresis angle, $\theta^\mathrm{eq}$. By extending the analysis in \cite{stoll1974analysis} to account for a nonzero hysteresis angle, the analytical eddy-current and hysteresis losses per unit area are given by
\begin{subequations}
	\label{eq:PePheq}
	\begin{align}
		&\label{eq:Peeq}P_\mathrm{e}=\frac{H_\mathrm{t}^2}{2\sigma\delta}\frac{1}{\cos\left(\frac{\theta^\mathrm{eq}}{2}\right)+\sin\left(\frac{\theta^\mathrm{eq}}{2}\right)},\\
		&\label{eq:Pheq}P_\mathrm{h}=\frac{H_\mathrm{t}^2}{2\sigma\delta}\frac{\sin{\theta^\mathrm{eq}}}{\cos\left(\frac{\theta^\mathrm{eq}}{2}\right)+\sin\left(\frac{\theta^\mathrm{eq}}{2}\right)},%
	\end{align}
\end{subequations}%
where $\delta = \sqrt{\frac{2}{\omega\sigma_{\mathrm{s}}\mu_0\mu_\mathrm{r}^\mathrm{eq}}}$ is the electromagnetic skin depth of the material. To determine the parameters $\mu_\mathrm{r}^\mathrm{eq}$ and $\theta^\mathrm{eq}$, the analytical linear losses in \eqref{eq:PePheq} are matched to the actual nonlinear losses. First, the nonlinear loss components $P_\mathrm{e}$ and $P_\mathrm{h}$ are computed as a function of $H_\mathrm{t}$ using a fully meshed finite-element model of the semi-infinite half-space. Within this model, the material's magnetic behavior is governed by the nonlinear complex permeability defined by either the EHL or FH framework, see Fig.~\ref{fig:murthetaS355}.

\begin{figure}[!t]
	\centering
	\includegraphics[width=\columnwidth]{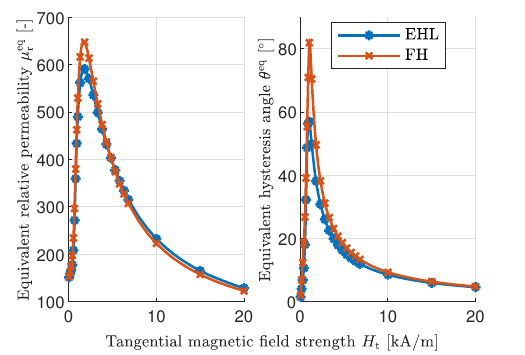}
	\caption{The equivalent permeability modulus and hysteresis angle as a function of the tangential magnetic field strength for both the EHL and FH method.} 
	\label{fig:Htmurtheta}
\end{figure}

Once these nonlinear losses are tabulated against $H_\mathrm{t}$, the equivalent parameters can be extracted analytically. Dividing \eqref{eq:Pheq} by \eqref{eq:Peeq} and taking the arcsine yields the equivalent hysteresis angle:
\begin{equation}
	\theta^\mathrm{eq}=\arcsin\left(\frac{P_\mathrm{h}}{P_\mathrm{e}}\right).
\end{equation}
Subsequently, by substituting the skin depth definition into \eqref{eq:Peeq}, the equivalent permeability modulus is obtained:
\begin{equation}
	\mu^\mathrm{eq}_\mathrm{r}=\frac{8\sigma P_\mathrm{e}P_\mathrm{tot}}{\omega\mu_0H_\mathrm{t}^4}.
\end{equation}
The resulting permeability modulus and hysteresis angle as a function of the applied surface field $H_\mathrm{t}$ are shown in Fig.~\ref{fig:Htmurtheta}. The decrease in the equivalent relative permeability and hysteresis angle at higher field strengths clearly indicates material saturation, matching the behavior shown in Fig.~\ref{fig:murthetaS355}

\subsection{Single-valued $B{-}H$ law with hysteresis loss curve}
A more traditional approach for estimating the steel's losses introduces hysteresis only after a field solution has been obtained. To find a solution for the field distribution, the $B{-}H$ relationship is treated as purely real and single-valued via the constitutive relationship in \eqref{eq:const_B}. The first method considered assumes a nonlinear $B{-}H$ constitutive law and uses volumetric meshing. The second method assumes a linear relationship via the use of SIBCs, and is therefore unmeshed.

\subsubsection{Nonlinear $B{-}H$ curve}
\begin{figure}[!t]
	\centering
	\includegraphics[width=\columnwidth]{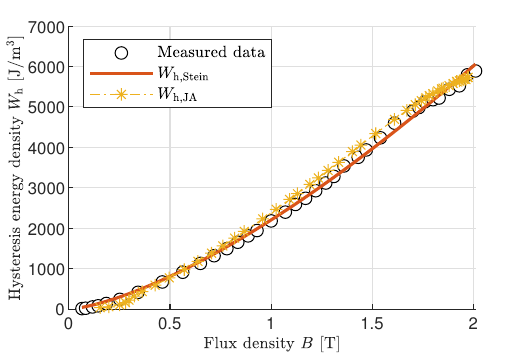}
	\caption{Hysteresis energy density versus maximum magnetic flux density for measured data \eqref{eq:Whys}, the Steinmetz equation \eqref{eq:stein}, and the Jiles-Atherton model.} 
	\label{fig:Wh_meas_Stein_JA}
\end{figure}

Within this framework, the magnetic field solution is governed by the nonlinear single-valued $B{-}H$ curve depicted in Fig.~\ref{fig:single_valued}. Once the field solution is obtained, hysteresis losses are calculated in a post-processing step. This step evaluates Steinmetz’s equation over the shell cross-section using the solved magnetic flux density. The resulting hysteresis energy density, in joules per unit volume, is defined as
\begin{equation}
	\label{eq:stein}
	W_\mathrm{h,Stein}(B) = k_\mathrm{h} B^{\alpha_\mathrm{s}}.
\end{equation}
The hysteresis loss scales linearly with the excitation frequency. The coefficients $k_\mathrm{h}$ and $\alpha_\mathrm{s}$ are determined by fitting \eqref{eq:stein} to the hysteresis energy density calculated in \eqref{eq:Whys}. Fig.~\ref{fig:Wh_meas_Stein_JA} plots the fit and measured data, highlighting the power-law relationship between magnetic flux density and hysteresis energy density. Table~\ref{tab:BHStein_coef} shows the numerical values.

\subsubsection{Linear $B{-}H$ curve for SIBCs}
This framework assumes a constant permeability throughout the material during solving, and uses post-processing to acquire the hysteresis loss. Therefore, the model uses SIBCs, eliminating the need to mesh the steel domain.

Since the SIBC replaces the meshed steel domain, the field solution does not contain the magnetic flux density throughout the steel. However, the flux density in linear materials follows a known exponential decay \cite{stoll1974analysis}. For a semi-infinite steel medium occupying $x>0$ as shown in Fig.~\ref{fig:Hs_interface}, the flux density is expressed as:
\begin{equation}
	B(x,H_\mathrm{t}) = B_\mathrm{s}(H_\mathrm{t})\exp{\left(-\frac{x}{\delta(\mu_\mathrm{r}(H_\mathrm{t}))}\right)},
	\label{eq:Blin}
\end{equation}
where the skin depth $\delta$, through its dependence on the relative permeability, depends on the tangential magnetic field strength. The relationship $\mu_\mathrm{r}(H_\mathrm{t})$ is equivalent to that of the EHL method in \eqref{eq:murBpHp} and Fig.~\ref{fig:murthetaS355} in which the horizontal axis is translated to $H_\mathrm{t}$. The flux density amplitude at $x=0$, $B_\mathrm{s}$, is mapped via $H_\mathrm{t}$ using the steels $B{-}H$ curve shown in Fig.~\ref{fig:single_valued}.

Substituting \eqref{eq:Blin} into \eqref{eq:stein} and integrating with respect to $x$ from $0$ to $\infty$ yields the hysteresis loss per unit area:
\begin{equation}
	P_\mathrm{h,post}(H_\mathrm{t})=k_\mathrm{h}f|B_\mathrm{s}(H_\mathrm{t})|^{\alpha_\mathrm{s}}\frac{\delta(\mu_\mathrm{r}(H_\mathrm{t}))}{\alpha_\mathrm{s}}.
\end{equation}
This post-processed hysteresis loss per unit area can then be integrated over the steel surface geometry to calculate the total hysteresis loss. The relative permeability for the S355 mild steel SIBC is set to $400$. This empirically derived value was chosen to minimize the absolute error against the experimental measurements presented later in this paper.

\subsection{Phenomenological hysteresis}
While several phenomenological frameworks exist for modeling hysteresis, notably the Preisach model \cite{mayergoyz1991}, the physics-based Jiles-Atherton model is particularly prevalent in practical applications \cite{jiles1986theory, szewczyk2018two}.

Assuming collinear applied field and magnetization ($M = \vert{}\vec{M}\vert{}$, $H = \vert{}\vec{H}\vert{}$), the magnetization $M$ in the JA constitutive relationship \eqref{eq:sommerfeld} can be described by an ordinary differential equation. Following \cite{szewczyk2018two, szewczyk2014computational, szewczyk2018open}, this ODE is given by:
\begin{equation}
	\label{eq:JAode}
	\frac{\mathrm{d}M}{\mathrm{d}H} = \frac{\delta_\mathrm{M}}{1 + c} \frac{M_\mathrm{an} - M}{k\delta_\pm - \alpha_\mathrm{c} (M_\mathrm{an} - M)} + \frac{c}{1 + c} \frac{\mathrm{d}M_\mathrm{an}}{\mathrm{d}H},
\end{equation}
where $c$ is the magnetization reversibility coefficient, $k$ is the pinning parameter, and $\alpha_\mathrm{c}$ represents the Bloch interdomain coupling. The directional parameter $\delta_\pm = \text{sign}\left(\frac{\mathrm{d}H}{\mathrm{d}t}\right)$, and $\delta_\mathrm{M} = \frac{1}{2} + \frac{1}{2}\text{sign}\left((M_\mathrm{an}-M)\frac{\mathrm{d}H}{\mathrm{d}t}\right)$ ensures that the incremental susceptibility remains strictly positive. This ODE is solved numerically to determine the magnetization increment $\mathrm{d}M$ for a given field increment $\mathrm{d}H$. The derivation of \eqref{eq:JAode} decomposes magnetization into reversible and irreversible components. Although often cited for its physical basis, this approach is merely postulated and may not reflect the actual underlying physical mechanisms \cite{zirka2012physical}.

The anhysteretic (hysteresis-free) magnetization $M_\mathrm{an}$ for isotropic materials is governed by the Langevin function:
\begin{equation}
	M_\mathrm{an} = M_\mathrm{s} \left( \coth\left( \frac{H_\mathrm{e}}{a} \right) - \frac{a}{H_\mathrm{e}} \right),
\end{equation}
where $M_\mathrm{s}$ is the saturation magnetization, $a$ characterizes the domain wall density, and the effective magnetic field strength is defined as $H_\mathrm{e} = H + \alpha_\mathrm{c} M$.

A challenge in applying the JA model is the inverse problem of accurately identifying its five parameters ($M_\mathrm{s}$, $a$, $\alpha_\mathrm{c}$, $k$, $c$) from measured data. As detailed in \cite{szewczyk2018two}, the five JA parameters can be determined using a differential evolution optimization algorithm. This approach is suitable because the inverse Jiles-Atherton modeling problem is characterized by numerous local minima. The objective function proposed is modified for permeable S355 steel instead of ferrite materials by assigning increased weight to the saturated regions of the hysteresis curve:
\begin{equation}
	F = \sum_{i}^{} w_i \left( B_\mathrm{meas}(H_i) - B_\mathrm{sym}(H_i) \right)^2,
\end{equation}
where $w_i=110$ if $B_\mathrm{meas}(H_i) > 0.7\,\max{(B_\mathrm{meas})}$ and $w_i =1$ otherwise. The ODE formulation in \eqref{eq:JAode} differs slightly from the implementation in COMSOL Multiphysics \cite{comsol_acdc_v64}. Consequently, Table~\ref{tab:JA_coef} (Appendix) provides the optimized parameters for both formulations for S355 steel.

For validation, hysteresis curves are reconstructed for all measured magnetic field excitations using the five parameters derived from the differential evolution algorithm. Two of these curves are shown in Fig.~\ref{fig:twoloops}. Numerically integrating the curves yields the hysteresis energy density $W_\mathrm{h,JA}$, shown as the third curve in Fig.~\ref{fig:Wh_meas_Stein_JA}.
\section{Experimental benchmark of the modeling frameworks}
\label{sec:experimentbenchmark}
This section outlines the experimental benchmark of the proposed loss frameworks, and uses the geometry as depicted in Fig.~\ref{fig:cylinder_geom}. The physical implementation of the setup is elaborated, and the predictions of all proposed frameworks are compared against experimental measurements.

\subsection{Design considerations}
\begin{figure}[!t]
	\centering
	\includegraphics[width=0.65\columnwidth]{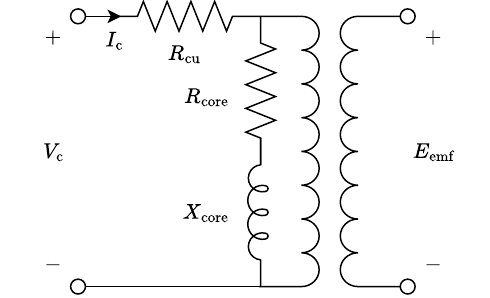}
	\caption{Equivalent circuit of the experimental setup.} 
	\label{fig:circuitsetup}
\end{figure}

\begin{figure}[!t]
	\centering
	\includegraphics[width=0.6\columnwidth]{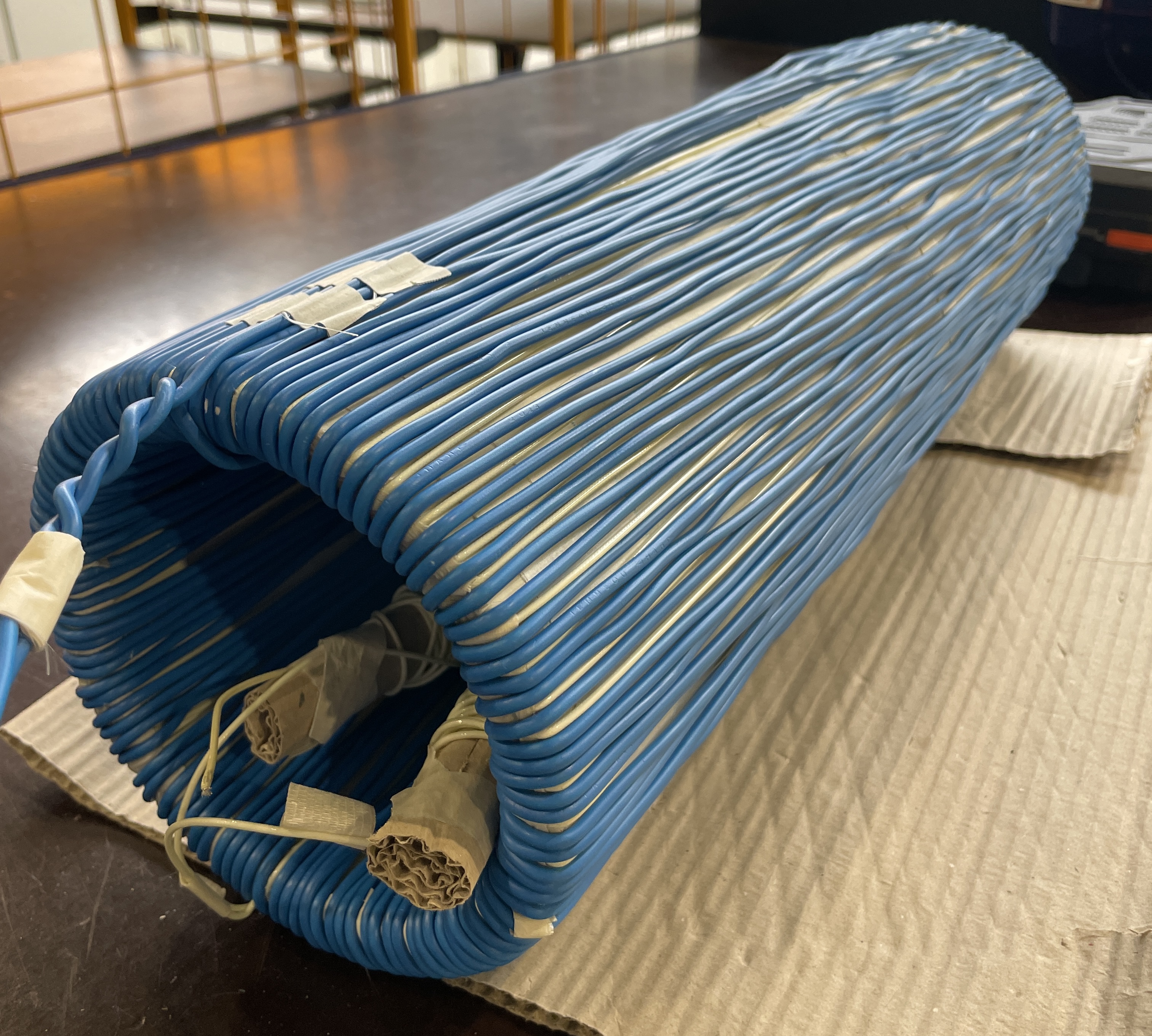}
	\caption{S355 steel cylindrical shell, sequentially layered from the surface outward with paint, Nomex transformer paper, a Nomex search coil, and a PVC-insulated copper conductor.} 
	\label{fig:shellsetup}
\end{figure}

The target magnetic flux density is set to $ 2\,\si{T}$ to encompass the full data range shown in Fig.~\ref{fig:murthetaS355}. Since the skin depth is around $2\,\si{mm}$, setting the shell thickness to $d = 1\,\si{cm}$ ensures that $d \gg \delta$. Additionally, $l=0.5\,\si{m}$ is chosen so that end-winding effects can be neglected. A $2.5\,\si{mm^2}$ PVC-insulated solid copper conductor was chosen to optimize the trade-off between ampacity and mechanical flexibility, while maintaining a compact overall diameter of $3.2\,\si{mm}$. 
An inner diameter of $132\,\si{mm}$ was selected. Although this diameter theoretically permits up to 130 turns, practical limitations associated with manual winding resulted in a final value of $N=107$. A 50-turn search coil was co-wound with the PVC-insulated copper conductor to directly measure the induced EMF, $E_\mathrm{emf}$. By isolating the EMF measurement from the coil voltage $V_\mathrm{c}$, the temperature-dependent ohmic drop of the copper resistance $R_\mathrm{cu}$ is bypassed (see Fig.~\ref{fig:circuitsetup}). Consequently, power is measured directly across the core impedance $R_\mathrm{core}+jX_\mathrm{core}$. The finalized setup is shown in Fig.~\ref{fig:shellsetup}. Excitation is provided by an ITECH 7900P power amplifier, and power measurements are acquired using a C.A 8335 power analyzer.
\subsection{Results}
This section compares the simulated losses of each framework against the experimentally measured losses. The models were implemented using the COMSOL Multiphysics 2D Magnetic Fields (mf) interface. Because the full geometry consists of 107 identical single-turn sectors, the domain is reduced to a single $3.36^{\circ}$ sector, as depicted in Fig.~\ref{fig:shell_mesh}.

\begin{table*}[!t]
	\caption{Summary of each framework full name, abbreviation as used in the loss plots and heatmap, study domain, mesh category, element count, and COMSOL computation time.}
	\centering
	\begin{tabular}{@{}llllll@{}}
		\toprule
		Framework name & Abbreviation & Study domain & Mesh & \# Mesh elements & Computation time (\si{s}) \\ \midrule
		Equivalent hysteresis losses & EHL & Frequency & Meshed & $976$ & $85.7$ \\
		Fundamental harmonics & FH & Frequency & Meshed & $976$ & $89.7$ \\
		Equivalent hysteresis losses for SIBCs & EHL SIBC & Frequency & Unmeshed & $226$ & $51.7$ \\
		Fundamental harmonics for SIBCs & FH SIBC & Frequency & Unmeshed & $226$ & $53.1$ \\
		Nonlinear $B{-}H$ curve & Nonlin. $B{-}H$ ($f$) & Frequency & Meshed & $976$ & $124.7$ \\
		Nonlinear $B{-}H$ curve & Nonlin. $B{-}H$ ($t$) & Time & Meshed & $976$ & $5052.4$ \\
		Linear $B{-}H$ curve for SIBCs & Lin. $B{-}H$ & Frequency & Unmeshed & $226$ & $48.8$ \\
		Jiles-Atherton & JA & Time & Meshed & $798$ & $23612$ \\
		\bottomrule
	\end{tabular}
	\label{tab:framework_comparison}
\end{table*}
\begin{figure*}[!t]
	\centering
	\includegraphics[width=\textwidth]{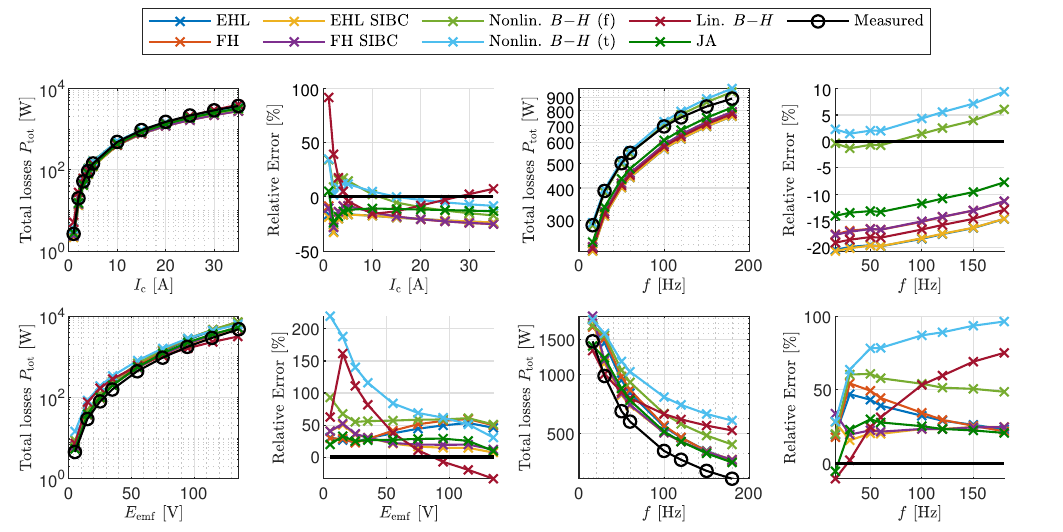}
	\caption{Losses and errors of all eight frameworks under current (top row) and voltage (bottom row) excitation. \textbf{Left:} Absolute total loss at $50\,\si{Hz}$. \textbf{Middle-left:} Relative error at $50\,\si{Hz}$. \textbf{Middle-right:} Absolute total loss versus frequency at $10\,\si{Arms}$ (top) and $65\,\si{Vrms}$ (bottom). \textbf{Right:} Relative error versus frequency at $10\,\si{Arms}$ (top) and $65\,\si{Vrms}$ (bottom).}
	\label{fig:losses_frameworks_shell_50Hz}
\end{figure*}
\begin{figure*}[!t]
	\centering
	\includegraphics[width=0.9\textwidth]{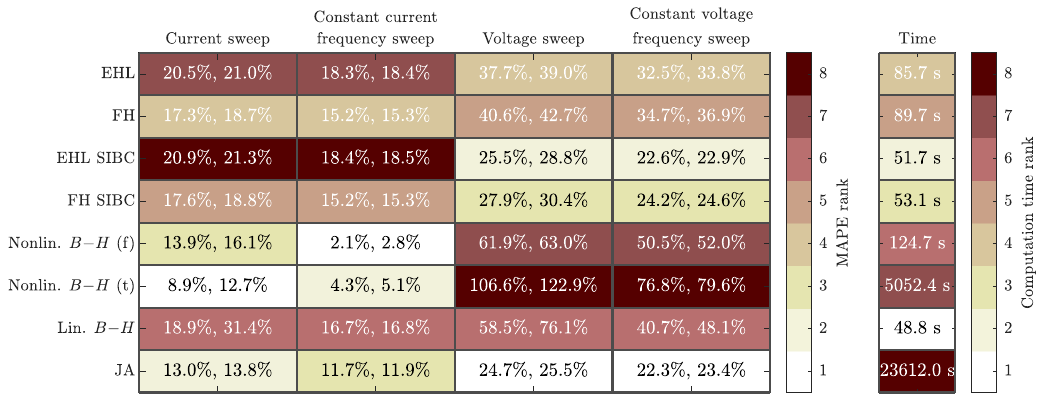}
	\caption{\textbf{Left:} Framework error metrics across four excitation types. Values within each cell represent MAPE (left) and RMSPE (right). Background shading indicates the relative MAPE ranking for that specific excitation type (column), fading from rank 1 (smallest error) to rank 8 (largest error). \textbf{Right:} Framework computation times. Background shading indicates the time ranking for each framework (row), fading from rank 1 (fastest) to rank 8 (slowest).}
	\label{fig:heatmap_shell_50Hz}
\end{figure*}
Section~\ref{sec:aclossshell} assumes linear magnetic properties for the shell. Under this linear assumption, a sinusoidal voltage excitation yields a purely sinusoidal current and vice versa. In other words, the equivalent core impedances $R_\mathrm{core}$ and $X_\mathrm{core}$ remain constant. This equivalence no longer holds for a physical nonlinear shell. In such cases, applying a purely sinusoidal excitation induces a distorted response. Consequently, the frameworks are benchmarked separately under sinusoidal current and voltage excitations. Note that while the ITECH features a built-in current source mode that maintains a set amplitude ($I_c$), its switch-mode nature combined with the highly nonlinear load results in a current THD of approximately $28\%$. COMSOL simulations indicate that this non-sinusoidal current waveform induces roughly $3\%$ additional total loss at high excitations.

Beyond these excitations, understanding the behavior of all frameworks under variable frequencies provides additional insight. Accordingly, losses were recorded experimentally and for all simulated frameworks at multiples of $50\,\si{Hz}$ and $60\,\si{Hz}$, and two additional low-frequency operating points.

Fig.~\ref{fig:losses_frameworks_shell_50Hz} provides an overview of the simulated and measured losses across all eight frameworks. The four subfigures on the left analyze the effects of varying the excitation amplitude at a constant frequency of $50\,\si{Hz}$. Specifically, the top-left subfigures display the total losses and relative errors for an increasing current excitation, while the bottom-left subfigures present the corresponding results for an increasing voltage excitation. These amplitude sweeps were also conducted at $60\,\si{Hz}$, with no observable differences in the resulting trends. The subfigures on the right depict the frequency-dependent behavior of the frameworks. The top-right subfigures show the total losses and relative errors across a frequency sweep under a constant current amplitude, whereas the bottom-right subfigures depict the corresponding results for a constant voltage amplitude. For each framework, the abbreviation as used in Fig.~\ref{fig:losses_frameworks_shell_50Hz} and Fig.~\ref{fig:heatmap_shell_50Hz}, study domain, mesh, number of mesh elements, and computation time to compute all losses are presented in Table~\ref{tab:framework_comparison}. To quantitatively evaluate the frameworks against the experimental measurements, the data from all subplots are further analyzed to extract key error metrics like the mean absolute percentage error (MAPE) and the root mean square percentage error (RMSPE). Percentage error is used instead of absolute error to prevent disproportionately weighting the high losses that occur at high current or voltage excitations.

Furthermore, comparing the MAPE and RMSPE provides insight into the error distribution of a given framework. Because RMSPE disproportionately penalizes larger percentage errors, a greater disparity between the two metrics indicates a wider dispersion of errors within the simulated losses. For clarity, the metrics are presented as a heatmap and shaded according to their MAPE rank, as shown in Fig.~\ref{fig:heatmap_shell_50Hz}.

Regarding the results, the unmeshed and frequency-domain frameworks require the least computation time, whereas the time-domain methods are slower due to time-stepping.

The sinusoidal current excitation includes the current sweep and constant-current frequency sweep (the top row in Fig.~\ref{fig:losses_frameworks_shell_50Hz}, and two leftmost columns in Fig.~\ref{fig:heatmap_shell_50Hz}). Under these excitations, the nonlinear $B{-}H$ frameworks in both domains yield the lowest MAPE and RMSPE, followed by the JA framework. Notably, the linear $B{-}H$ framework using SIBCs exhibits the largest discrepancy between its MAPE and RMSPE, driven by outliers at low currents.

The sinusoidal voltage excitation includes the voltage sweep and constant-voltage frequency sweep (the bottom row in Fig.~\ref{fig:losses_frameworks_shell_50Hz}, and two rightmost columns in Fig.~\ref{fig:heatmap_shell_50Hz}). The JA and SIBC-based complex permeability frameworks achieve the lowest MAPE, altough the MAPE values generally increase compared to sinusoidal current excitation. Notably, the three $B{-}H$ frameworks yield relatively high MAPE values because their field solutions neglect hysteresis loss. For this reason, these models simulate an overall shell impedance that is roughly $20\%$ lower at peak MAPE compared to the JA framework for the same EMF. For a set EMF, this artificially low impedance simulates an erroneously high coil current, directly causing an overestimation of eddy-current loss. On top of this and as demonstrated by the linear case in Section~\ref{sec:aclossshell} (Fig.~\ref{fig:cylinder_loss}), hysteresis inherently reduces the eddy-current loss. Because post-processing methods ignore hysteretic damping on the current density, they fail to account for this corresponding reduction. In essence, these frameworks add hysteresis loss to an undamped eddy-current loss. Therefore, incorporating hysteresis loss is essential for accurate loss estimation, rather than adding it during post-processing or neglecting it altogether

\section{Case Study}
\label{sec:casestudy}
\begin{figure}[!t]
	\centering
	\subfloat[CAD render.]{\includegraphics[width=0.75\columnwidth]{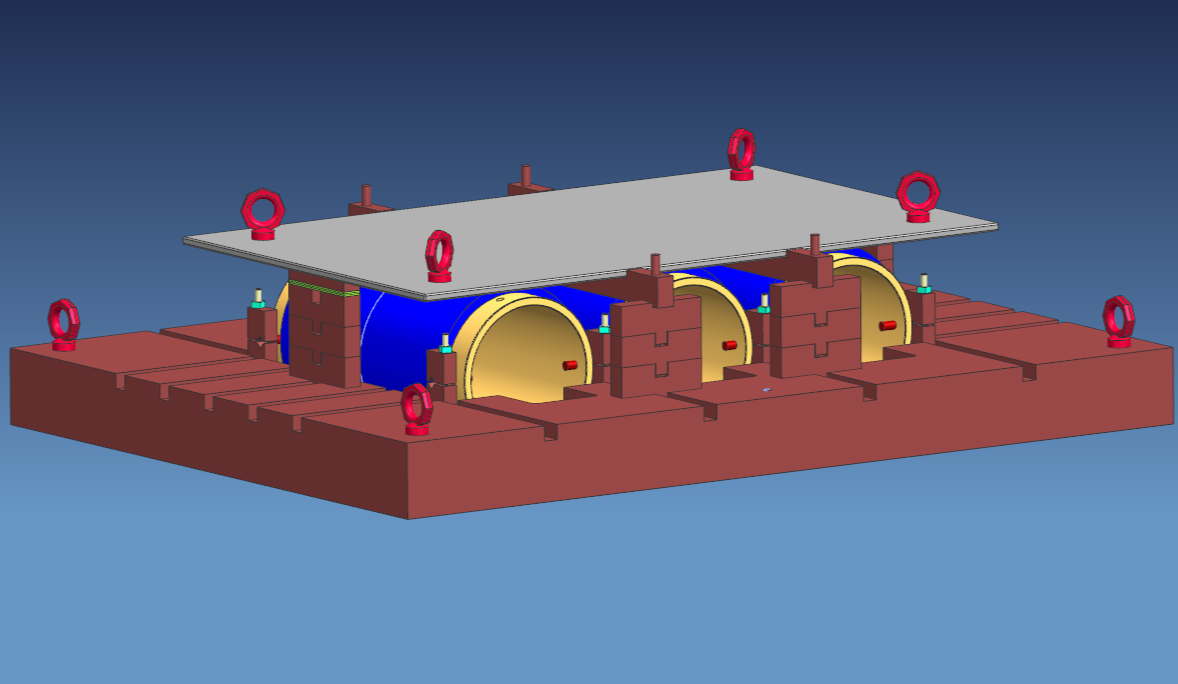}%
		\label{fig:reactor_render}}
	\\
	\subfloat[Physical setup.]{\includegraphics[width=0.75\columnwidth]{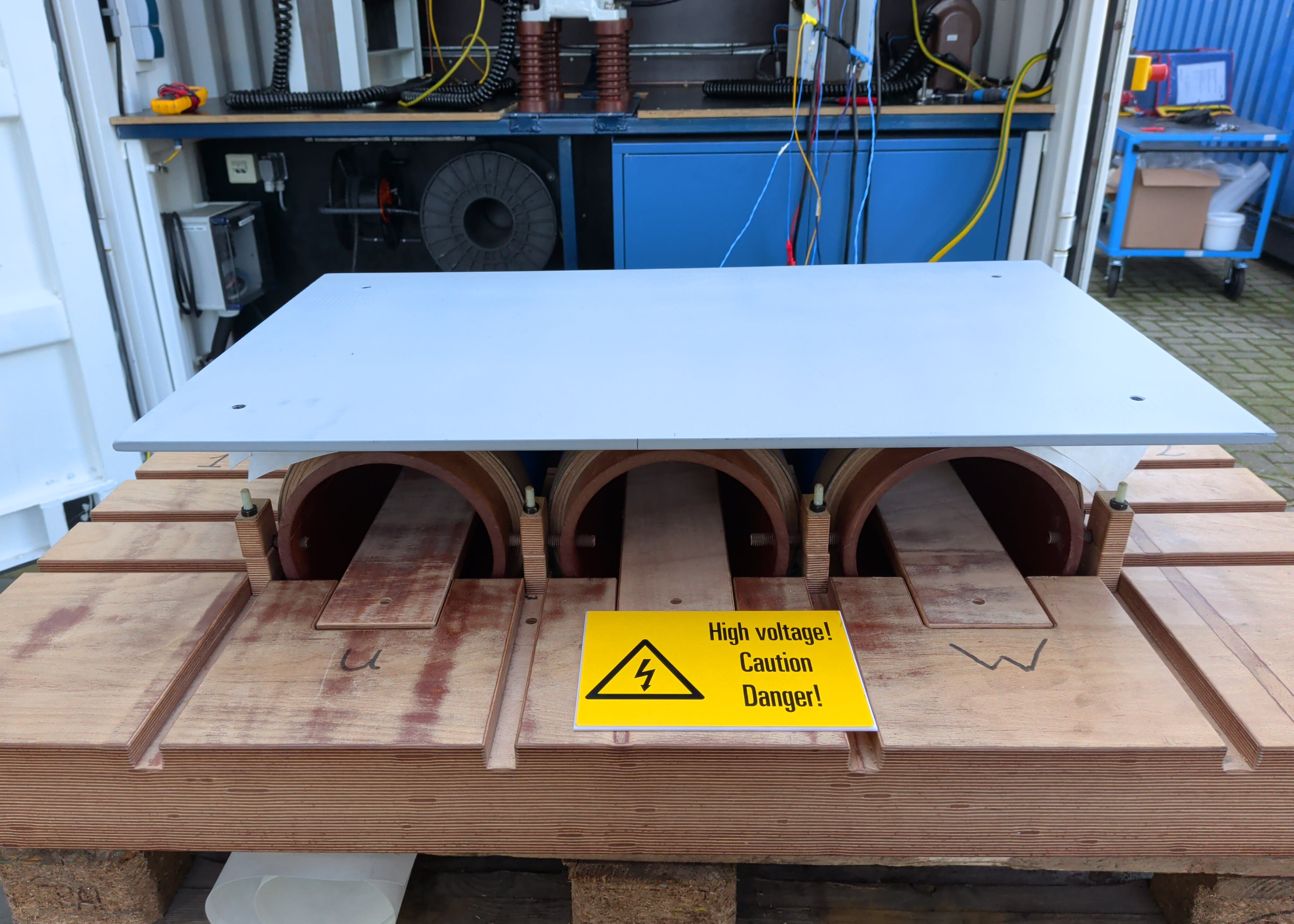}%
		\label{fig:reactor_setup}}
	\caption{Geometry of the reactor setup: wooden support at the base, three horizontally mounted wooden bobbins with copper windings and search coil, and S355 mild steel tank wall at the top. (a) CAD render. (b) Physical setup.} 
	\label{fig:combined_setup}
\end{figure}

To analyze how the frameworks behave in a power transformer geometry, a scale-model, three-phase reactor is considered to compute losses induced in a mild steel tank wall. The reactor has no magnetic core, is connected in star, and is mounted on a wooden support. The tank wall is centered above the middle winding, see Fig.~\ref{fig:combined_setup}. Three concentric search coils are wrapped around the outer circumference of each four-layer primary winding to perform EMF measurements. All geometric parameters are listed in Table~\ref{tab:reactor_parameters}. Unlike the cylindrical shell in Section~\ref{sec:experimentbenchmark}, the impedance of the reactor setup is linear, resulting in no distinction between sinusoidal voltage and current excitation.

In order to obtain the total tank wall losses experimentally, measurements are performed on the search coils and primary windings. By substituting the scaled search coil voltage for the primary voltage, the active power transferred to the tank wall is evaluated while bypassing the primary ohmic losses:
\begin{equation}
	P_{\mathrm{wall}} = k_{\mathrm{g}} \frac{N_{\mathrm{p}}}{N_{\mathrm{s}}} \sum_{i=1}^{3} \Re\left(V_{\mathrm{s},i} I_{\mathrm{p},i}^* \right)
	\label{eq:Pwall}
\end{equation}
where $V_{\mathrm{s},i}$ and $I_{\mathrm{p},i}$ are the rms search coil voltage and primary current phasors of phase $i$, $N_{\mathrm{s}}$ and $N_{\mathrm{p}}$ are the number of turns in the search coil and primary winding, respectively, and $k_{\mathrm{g}}$ is a lumped geometric correction factor. 

The factor $k_{\mathrm{g}}$ accounts for two geometric phenomena that cause the physical flux linkage to deviate from the ideal turns ratio. First, since the search coil is positioned at the outer radius of the primary winding, it encloses flux that the inner primary layers miss. Second, when the conductive tank wall is introduced, the spatial decay of its out-of-phase reaction field links the outer search coil more strongly than the inner primary coil. Consequently, relying solely on the nominal turns ratio overestimates the transferred active power. Calibrating \eqref{eq:Pwall} against the direct volumetric integration of the tank wall losses in an FEA model incorporates both geometric effects into a single correction factor of $k_{\mathrm{g}} = 0.87$. Proximity and skin-effect stray losses within the primary turns increase the winding total loss by only $1.64\%$ over its DC value and therefore have a negligible impact.
\begin{figure*}[!t]
	\centering
	\includegraphics[width=8cm]{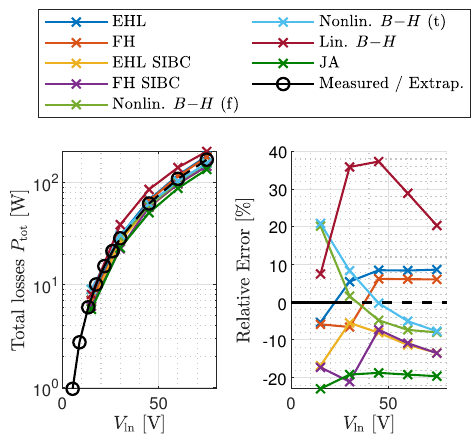}%
	\includegraphics[width=10cm]{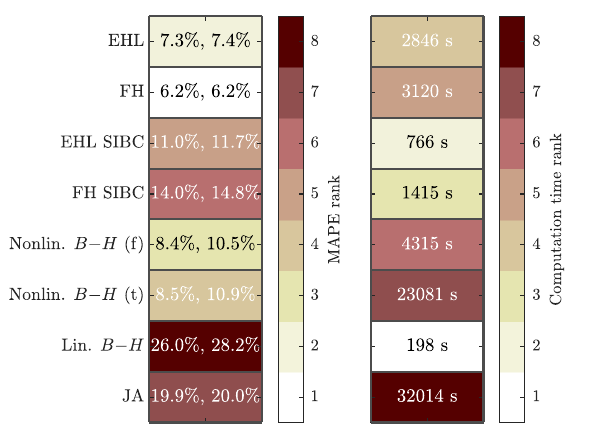}
	\caption{Losses and errors of all eight frameworks under voltage excitation at $50\,\si{Hz}$. \textbf{Left:} Absolute total loss. \textbf{Middle-left:} Relative error. \textbf{Middle-right:} Framework error metrics. Values within each cell represent MAPE (left) and RMSPE (right). \textbf{Right:} Framework computation times.}
	\label{fig:Losses_aircore}
\end{figure*}

Fig.~\ref{fig:Losses_aircore} compares the measured and simulated total losses across a range of line-to-neutral voltages, $V_\mathrm{ln}$, as well as their fastest computation time. The absolute losses for the measurements and all eight frameworks are shown on the left, while the corresponding relative errors, which generally remain within $\pm20\%$, are displayed on the right. Excitation was provided via a Lechmotoren SDV 49.41-4 synchronous generator, and losses were measured at six voltage levels using a Newtons4th PPA5530-30A precision power analyzer. Although this analyzer maintains high accuracy at low power factors (a phase accuracy of $0.0055^{\circ}$ at $50\,\si{Hz}$), its maximum current rating is limited to $30\,\si{A}$. To estimate losses at higher flux densities in the plate, the measured data was extrapolated using a power-law fit, yielding $P_{\mathrm{wall}} = 0.04042 V_\mathrm{ln}^{1.9285}$ ($R^2 = 0.9998$). Consistent with Section~\ref{sec:experimentbenchmark}, frameworks that post-process hysteresis loss yield a noticeably larger error at lower excitations (where the hysteresis contribution is greater) compared to higher excitations. While consistent, the JA framework yields a relatively high error, which can be attributed to batch-to-batch material variations in the mild steel

\begin{table}[!t]
	\caption{Geometric parameters of the reactor setup, which are also used in FEA simulations.}
	\centering
	\begin{tabular}{@{}llll@{}}
		\toprule
		Quantity                               & Symbol                 & Value               & Unit      \\ \midrule
		Winding inner radius                   & $r_{\mathrm{w,in}}$    & $101$               & \si{mm}   \\
		Winding outer radius                   & $r_{\mathrm{w,out}}$   & $114$               & \si{mm}   \\
		Winding height                         & $h_{\mathrm{w}}$       & $32$                & \si{cm}   \\
		Winding fill factor                    & $F$                    & $0.85$              & --        \\
		Tank wall height                       & $h_{\mathrm{t}}$       & $55$                & \si{cm}   \\
		Tank wall width                        & $w_{\mathrm{t}}$       & $90$                & \si{cm}   \\
		Tank wall thickness                    & $d_{\mathrm{t}}$       & $1$                 & \si{cm}   \\
		Center-to-center winding distance      & $d_{\mathrm{cc}}$      & $25$                & \si{cm}   \\
		Minimum phase-to-plate air gap         & $g$                    & $1$                 & \si{cm}   \\
		Number of turns per phase              & $N_{\mathrm{p}}$       & $160$               & --        \\
		Number of turns per search coil        & $N_{\mathrm{s}}$       & $169$               & --        \\
		\bottomrule
	\end{tabular}
	\label{tab:reactor_parameters}
\end{table}

\section{Conclusion}
In summary, this paper presented a comparative analysis of eight frameworks for predicting eddy-current and hysteresis losses in bulky mild steel, evaluating the trade-off between accuracy and computational cost. By deriving the electromagnetic properties of S355 steel from experimental measurements, the parameters for each framework were tuned to the steel using identification procedures ranging from standard curve-fitting and Fourier transforms to a differential evolution algorithm. Under current excitation, the loss frameworks for the cylindrical magnetic shell achieved mean absolute percentage errors below $20\%$, with all but the linear $B{-}H$ curve model yielding consistent results. Under voltage excitation, errors remained below $40\%$. However, the three frameworks computing hysteresis loss via post-processing produced larger errors because shell impedance is coupled to the hysteresis loss within the field solution. Additionally, harmonic distortion in the experimental current excitation caused slightly higher measured losses. For the reactor setup, the relative error stayed generally within $\pm20\%$. Both setups demonstrate the necessity of incorporating the hysteresis contribution directly into the field solution, rather than post-processing or neglecting it. Furthermore, a comparison of computation times shows that for large magnetic devices, SIBC-based frameworks offer an effective and computationally efficient approach. Future work could extend this benchmarking to diverse transformer geometries while specifically including the hysteresis component. Moreover, they could identify the electromagnetic properties of other structural steels and quantify their batch-to-batch variation.

\appendices
\section{Parameters of the single-valued $B{-}H$ curve, Steinmetz equation and JA model}
\begin{table}[h]
	\caption{Coefficients obtained during fitting of the single-valued $B{-}H$ curve in \eqref{eq:BHfitted} and Steinmetz's equation in \eqref{eq:stein}.}
	\centering
	\begin{tabular}{@{}llll@{}}
		\toprule
		Quantity & Symbol & Value & Unit \\ \midrule
		\multirow{7}{*}{\begin{tabular}[c]{@{}l@{}}Single-valued $B{-}H$\\ curve coefficients\end{tabular}} 
		& $c_1$      & $0.4476 $ & \si{Tm/A}\\
		& $c_2$      & $0.2357 $ & \si{Tm^2/A^2}\\
		& $c_3$      & $0.0329 $ & \si{m^3/A^3}\\
		& $c_4$      & $4.1789\times 10^7 $ & $-$ \\
		& $c_5$      & $-4.4166\times 10^4 $ & \si{m/A} \\
		& $c_6$      & $43.3555 $ & \si{m^2/A^2}\\
		& $B_\mathrm{max}$  & $2 $ & \si{T} \\ \midrule
		\multirow{2}{*}{Steinmetz coefficients}
		& $k_\mathrm{h}$      & $2206$ & \si{J/(m^3T^{1.447})}\\
		& $\alpha_\mathrm{s}$ & $1.447$ & $-$ \\ \bottomrule
	\end{tabular}
	\label{tab:BHStein_coef}
\end{table}

\begin{table}[h]
	\caption{JA model parameters for S355 steel. The left column presents values for the formulation in \eqref{eq:JAode}, while the right column lists those for COMSOL Multiphysics.}
	\centering
	\begin{tabular}{@{}lllll@{}}
		\toprule
		Quantity            & Symbol    & Value  & Value & Unit     \\ \midrule
		Saturation magnetization   & $M_\mathrm{s}$    & $1.54$  & $1.54$ &   \si{MA/m}   \\
		Domain wall density   & $a$    & $980$  & $980$ &   \si{A/m}   \\
		Bloch interdomain coupling   & $\alpha_\mathrm{c}$    & $0.002$  & $0.002$ &   \si{-}   \\
		Pinning parameter   & $k$    & $690$  & $1000$ &   \si{A/m}   \\
		Reversibility coefficient   & $c$    & $0.67$  & $0.25$ &   \si{-}   \\
		\bottomrule
	\end{tabular}
	\label{tab:JA_coef}
\end{table}

\section*{Acknowledgment}
The research leading to these results was supported by Royal SMIT Transformers B.V., KTH Royal Institute of Technology, and Eindhoven University of Technology as part of the EIT InnoEnergy program Smart Electrical Networks and Systems (SENSE). This manuscript is derived from the first author's master's thesis, which introduced the initial methodology and findings regarding stray losses in power transformers \cite{swaanen2026stray}. Additionally, the authors acknowledge the use of generative AI for grammar review, code debugging, and LaTeX equation typesetting.

\bibliographystyle{IEEEtran}
\bibliography{IEEEabrv,ref}
\balance

\end{document}